\documentclass[10pt, conference, letterpaper]{IEEEtran}
\usepackage{cite} 
\usepackage{amsmath}
\usepackage{graphicx}
\usepackage{balance}
\usepackage{multirow}

\ifCLASSINFOpdf
\else
\fi

\author{
    \IEEEauthorblockN{Somayeh Kafaie\IEEEauthorrefmark{1}\IEEEauthorrefmark{2}, Mohamed H. Ahmed\IEEEauthorrefmark{2}, Yuanzhu Chen\IEEEauthorrefmark{1}\IEEEauthorrefmark{3},  Octavia A. Dobre\IEEEauthorrefmark{2}}
    \\\
    \IEEEauthorblockA{\IEEEauthorrefmark{1}Wireless Networking and Mobile Computing Laboratory}
    \IEEEauthorblockA{\IEEEauthorrefmark{2}Faculty of Engineering and Applied Science, Memorial University, St. John's, NL A1B 3X5, Canada}
    \IEEEauthorblockA{\IEEEauthorrefmark{3}Department of Computer Science, Memorial University, St. John's, NL A1B 3X5, Canada}
}

\usepackage{enumitem}
\begin{document}
%
\title{Throughput Analysis of Network Coding \\in Multi-Hop Wireless Mesh Networks \\Using Queueing Theory}



%


\maketitle

\begin{abstract}
In recent years, a significant amount of research has been conducted to explore the benefits of network coding in different scenarios, from both theoretical and simulation perspectives. In this paper, we utilize queueing theory to propose an analytical framework for bidirectional unicast flows in multi-hop wireless mesh networks, and study throughput of inter-flow network coding. We analytically determine performance metrics such as the probability of successful transmission in terms of collision probability, and feedback mechanism and retransmission. Regarding the coding process, our model uses a multi-class queueing network where coded packets are separated from native packets and have a non-preemptive higher priority over native packets, and both queues are in a stable state. Finally, we use simulations to verify the accuracy of our analytical model. 
\end{abstract}



%
\IEEEpeerreviewmaketitle

\section{Introduction}

Network coding, introduced by Ahlswede et al.~\cite{NC-Ahlswede-IEEETransactionsIT2000}, is an innovative idea to increase the transmission capacity of the network as well as its robustness. 
A number of studies have been conducted to investigate the benefits of network coding over the traditional forwarding approach in different scenarios. 
Most previous theoretical studies are valid only for saturated queues, where each node always has a packet to transmit that would cause an infinite delay. In addition, in many cases researchers model network coding with simplifying assumptions such as no interference among nodes, and no collision. In particular, to avoid collision in non-saturated queues, Sagduyu et al.~\cite{AnaNCQ-Sagduyu-IT2008} consider a conflict-free scheduled access.

There has been prior research done on multi-hop wireless networks as well. In particular, Johnson et al.~\cite{AnaIntraNCQ-Johnson-NetCod2011} study the performance of intra-flow network coding. Some other research investigates the throughput gain of inter-flow network coding over a non-coding scheme on multicast sessions~\cite{AnaNCQ-Sagduyu-IT2008, AnaNCQ-Iraji-WCNC2009, AnaNCQ-Amerimehr-MC2014}. Hwang et al.~\cite{AnaNCQunicast-Hwang-Net2011} investigate the performance of unicast sessions in multi-hop wireless networks, but they focus mostly on collision and interference levels, and do not provide any theoretical analysis for coding probability in relays.

In this paper, we provide an analytical framework based on multi-class queueing network to study the throughput of inter-flow network coding in multi-hop wireless mesh networks. 
Our model is provided for a multi-hop chain topology with bidirectional unicast flows in opposite directions, where intermediate nodes can combine the packets of two flows using \emph{XOR}. We model the packet forwarding process via network coding without postponing transmission of native packets to generate coded packets. Also, we consider separate classes of queues for native and coded packets, while the coded queue is the first-priority queue. 

The rest of this paper is organized as follows. Related work is discussed in Section \ref{relatedWork}. We explain the system model and assumptions in Section \ref{systemOverview}. Section \ref{problemFormulation} introduces our derived formulation of throughput in different scenarios. In order to show the accuracy of our analytical model, we compare the result with computer simulation in Section \ref{performanceEvaluation}. Finally, Section \ref{conclusion} draws conclusions, and discusses directions for future work. 

\section{Background and Related Work} \label{relatedWork}

There has been prior research on network coding from theoretical point of view, which mostly focuses on unicast sessions in a two-hop region, where a single relay forwards packets of multiple sources to their final destination~\cite{AnaNC-basic-Amerimehr-WC2014, Ana-NC-2hop-Zeng-PDS2014, Ana-NC-2hop-Le-MC2010,Ana-NC-2hop-Paschos-inforcom2013}. In addition, Khan et al.~\cite{Inter+Intra+PerformanceTwoSources-Khan-ICC2015} study the performance of the joint inter-flow and intra-flow network coding approach in a two-hop region, where two sources transmit packets to one destination via a relay. They calculate an upper bound on the probability that the destination recovers the packets of both sources successfully.

Sagduyu et al.~\cite{AnaNCQ-Sagduyu-IT2008} consider a collision-free scheduled access to formulate throughput for both saturated and non-saturated queues in multi-hop wireless networks. However, in case of random access scheme, their analytical model is limited to saturated queues. In this paper, instead of limiting nodes to scheduled access, we study the performance using IEEE 802.11 MAC layer, where collision can occur without assuming saturated queues. In addition, we provide simulation results to verify our model.

In a similar theoretical-based approach for multicast sessions, Amerimehr et al.~\cite{AnaNCQ-Amerimehr-MC2014} derive throughput for multi-hop wireless networks. However, in the coding process, they assume the decoding probability is equal to one and also they postpone transmission of the native packet at a node until receiving a packet from another flow to be combined with the first packet, and thus, causing a long delay. 

Furthermore, Hwang et al.~\cite{AnaNCQunicast-Hwang-Net2011} propose an analytical framework for bidirectional unicast flows in multi-hop wireless networks. Their work considers collision and different interference levels in CSMA/CA (Carrier Sense Multiple Access with Collision Avoidance) by varying carrier-sensing range and signal-to-interference ratio to maximize the throughput in different retransmission schemes. Although their scenario is similar to ours, our work is different in the following aspects: 1) in contrast to their approach, in our model, if a node has a transmission opportunity, it does not delay forwarding native packets to generate coded packets; 2) their focus is on saturated queues, while we work on stable queues.

\section{System Overview} \label{systemOverview}
\subsection{Network Model and Assumptions}

\begin{figure}[ht]
\centering
\includegraphics[scale=0.65]{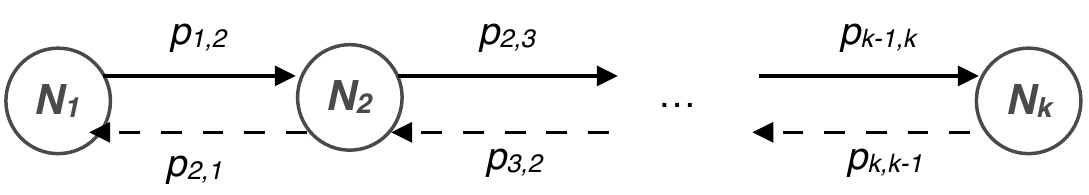}
\caption{Chain topology used for the analytical model.}
\label{chain}
\end{figure}

Our analytical model is proposed for UDP flows in a chain topology, as depicted in \figurename~\ref{chain}. In this network, there exist $k$ nodes, namely $N_{1}$ to $N_{k}$ with unlimited queue capacity. $N_{1}$ and $N_{k}$ transmit their packets to each other via the intermediate nodes $N_{2}$ to $N_{k-1}$. The model that we consider for interference assumes that nodes can not transmit and receive at the same time, and all transmissions in the range of the receiver are considered as interference. Furthermore, we assume that the feedback channel is reliable; thus if a node does not hear an ACK on time, it assumes that the data packet is lost.

We consider each node as a queuing system, where the packets in the sending buffer are customers of the queue and the node acts as the server. Hence, \emph{Queueing Theory} can be used to model this network. We assume that the queues are in a stable state, i.e., the arrival rate is less than the service rate.





\subsection{The Probability of Successful Transmission}\label{routing}
\begin{figure}[ht]
\centering
\includegraphics[scale=0.6]{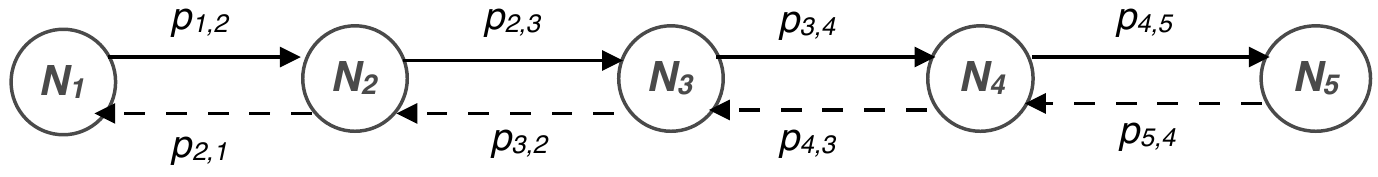}
\caption{Chain topology with 5 nodes.}
\label{chain-5}
\end{figure}

We assume that the probability that node $N_{i}$ successfully transmits a packet to its neighbor $N_{j}$ is $p_{i,j} > 0$, and equals zero for other nodes. This probability for each link is calculated in terms of collision. We assume that the probability of collision between a data packet and an ACK is negligible; this is a valid assumption because 1) the length of ACKs is significantly shorter than the length of data packets, and 2) ACKs are given higher priority and are sent earlier than any data packet. Therefore, in a chain topology with $5$ nodes as depicted in \figurename~\ref{chain-5}, a transmission from $N_{2}$ to $N_{3}$ will fail if at the same time slot that $N_{2}$ is transmitting, $N_{3}$ or $N_{4}$ transmits as well. Note that we assume that due to the capture effect, a transmission from $N_{1}$ or $N_{5}$ will not collide with reception at $N_{3}$. 

If $N_{2}$ transmits a packet at time $t$, after the propagation delay ($\delta$), $N_{3}$ and $N_{4}$ will sense that the channel is busy and avoid any transmission (assuming perfect time synchronization among nodes). Therefore, during a propagation delay window before and after $N_{2}$'s transmission (i.e., $(t-\delta, t+\delta)$), $N_{3}$ and $N_{4}$ may transmit their packet which will collide with $N_{2}$'s transmission. Although the propagation delay depends on the distance, we assume a fixed propagation delay as the maximum propagation delay.

The probability that $N_{3}$ transmits a packet during this time window equals $2 \delta \lambda_{3}$, and the probability that $N_{4}$ transmits a packet during the same time window is $2 \delta \lambda_{4}$. Therefore, the probability of a successful transmission from $N_{2}$ to $N_{3}$ equals $p_{2,3} = (1 - 2 \delta \lambda_{3}) (1 - 2 \delta \lambda_{4})$. In fact, the following equation can be used to compute the probability of successful transmission from $N_{i}$ to $N_{j}$, when $N_{i}$ and $N_{j}$ are neighbors:
\begin{equation}
p_{i,j}=\prod \limits_{x=m_{0}}^{m_{M}} (1-2\delta\lambda_{x})\, ,
\end{equation} 
where $M$ is the number of neighbors of $N_{j}$, and $m_{z}$ $( 0 \leq z \leq M, m_{z} \neq i)$ is one of the neighbors of $N_{j}$ or itself.

\begin{table}[!t]
\caption{The description of used symbols.}
\label{table:symbols}
\centering
\begin{tabular}{|l|l|}
\hline
\textbf{Symbol} & \textbf{Description}\\  \hline
$p_{i,j}$ & the probability of successful transmission from $N_{i}$ to $N_{j}$\\ \hline
$\delta$ & the maximum propagation delay \\ \hline
$\gamma_{i}$ & input rate at the source $N_{i}$	\\ \hline
$\lambda_{i}$ & the arrival rate at $N_{i}$  \\ \hline
$\mu$ & the service rate of the queue \\ \hline 
$\beta$ & the maximum number of transmissions of a packet \\ \hline
$\theta$ & throughput \\ \hline
$\lambda_{in,i}^{n(j)}$ &  the arrival rate of native packets of the $j^{th}$ flow at $N_{i}$\\ \hline
$\lambda_{out,i}^{n(j)}$ &  the output rate of native packets of the $j^{th}$ flow at $N_{i}$ \\ \hline
$\lambda_{in,i}^{c(j)}$ &  the arrival rate of coded packets of the $j^{th}$ flow at $N_{i}$\\ \hline
$\lambda_{out,i}^{c}$ &  the output rate of coded packets at $N_{i}$ \\ \hline
$\lambda_{i}^{n(j)}$ &  the arrival rate of the $j^{th}$ flow in the native queue of $N_{i}$\\ \hline
$\lambda_{i}^{c}$ &  the arrival rate in the coded queue of $N_{i}$	\\ \hline
$P_\text{mix}$ & the probability of combining two flows \\ \hline
 \end{tabular}
\end{table}

\section{Problem Formulation} \label{problemFormulation}
To construct our analytical model for network coding progressively, we propose it in several steps. We start with the basic case of one flow and no coding opportunity, where nodes do not retransmit any packet even if it is lost. Finally we build our model for two bidirectional flows between $N_{1}$ and $N_{k}$, where intermediate nodes may combine packets of the two flows, and the sender retransmits the packet if it does not hear any ACK. Table~\ref{table:symbols} presents all the variables used in this Section.


\subsection{Step 1- One Flow, no Retransmission, no Coding}
To start with a simple case, we model a network with one flow, no retransmission and no coding where non-coded packets travel from $N_{1}$ to $N_{k}$ in \figurename~\ref{chain}, and the sender will not retransmit any packet even if it is lost. This model is called \emph{Tandem queue} as the nodes form a series system with a single flow from the source to the destination, while the customers (i.e., packets) may enter from outside only at node $N_{1}$ and depart only from node $N_{k}$. We assume that there is no restriction on waiting between nodes. Thus, each node can be analyzed separately as a single $M/M/1/\infty$ queuing model, and the arrival rate in each node can be computed as:

\begin{equation}
\label{eq:lambda}
\lambda_{i} =
\left\{
	\begin{array}{ll}
		\gamma_{i}  & \mbox{if } N_{i}  \mbox{ is the source }  (i=1)\\
		\lambda_{i-1} \times p_{i-1,i}  & \mbox{if } 1 < i \leq k
	\end{array}
\right.
\end{equation}

\subsubsection{Throughput}
The throughput, denoted by $\theta$, is identical to the arrival rate at the destination $N_{k}$, which is presented by:
\begin{equation}
\theta=\lambda_{k}= \gamma_{1} \prod \limits_{i=1}^{k-1} p_{i,i+1}
\end{equation}

\subsubsection{Successful transmission probabilities}
As explained in Subsection~\ref{routing}, the probability of successful transmission within different links in our chain topology can be calculated by solving the following system of non-linear equations:
\begin{equation}
\label{eq:probabilities1-1}
\begin{cases} 
p_{1,2}=(1-2 \delta \lambda_{2})(1-2 \delta \lambda_{3}) \\  
...\\
p_{i-1,i}=(1-2 \delta \lambda_{i})(1-2 \delta\lambda_{i+1})\\
...\\
p_{k-2,k-1}=(1-2 \delta\lambda_{k-1})\\
p_{k-1,k}=1
\end{cases}
\end{equation}

Note that in this scenario $N_{k}$ does not send any data packet; hence, no interference from this node affects calculation of the successful transmission probabilities of the links. Recall that the arrival rate in each node depends on the arrival rate in the previous hop and the probability of successful transmission from previous hop to this node. If we calculate the arrival rates in different nodes of the route recursively, one may notice that all arrival rates can be computed in terms of the input rate at the source, $\gamma_{1}$, and successful transmission probabilities. Therefore, the system of non-linear equations described in~(\ref{eq:probabilities1-1}), for the topology depicted in \figurename~\ref{chain-5} (i.e., $k=5$), can be rewritten as:
\begin{equation}
\begin{cases} 
p_{1,2}=(1-2 \delta \gamma_{1} p_{1,2})(1-2 \delta \gamma_{1} p_{1,2}p_{2,3}) \\ 
p_{2,3}=(1-2 \delta \gamma_{1} p_{1,2}p_{2,3})(1- 2 \delta \gamma_{1} p_{1,2}p_{2,3}p_{3,4}) \\ 
p_{3,4}=(1-2 \delta \gamma_{1} p_{1,2}p_{2,3}p_{3,4}) \\
p_{4,5}=1 
\end{cases} 
\end{equation}

\subsection{Step 2- Two Flows, no Retransmission, no Coding} \label{step2}
In the second scenario, another flow is added in the opposite direction. Thus, two flows are initiated from $N_{1}$ and $N_{k}$, while the intermediate nodes forward only native packets, and the sender will not retransmit any packet even if it is lost. In this scenario, the packets may enter the network (i.e., the queue network) either at node $N_{1}$ with an arrival rate $\gamma_{1}$ or at node $N_{k}$ with an arrival rate $\gamma_{k}$, and depart from the other end of the chain. Therefore, the intermediate nodes receive the packets from both directions.


Let $\lambda_{i}^{(1)}$ and $\lambda_{i}^{(2)}$ denote the arrival rate of the first flow (i.e., from $N_{1}$ to $N_{k}$) and the second flow (i.e., from $N_{k}$ to $N_{1}$) arriving at node $N_{i}$, where $\lambda_{1}^{(1)}=\gamma_{1}$ and $\lambda_{k}^{(2)}=\gamma_{k}$. Therefore, at each node $\lambda_{i}=\lambda_{i}^{(1)} + \lambda_{i}^{(2)}$. Eq.~(\ref{eq:lambda2}) presents the arrival rate of packets of the two flows at different nodes.
\begin{equation}
\label{eq:lambda2}
\left\{
	\begin{array}{ll}
	   \lambda_{i}^{(1)}=\gamma_{i}   & \mbox{if } i=1\\
		\lambda_{i}^{(1)}= \lambda_{i-1}^{(1)} \times p_{i-1,i} & \mbox{if } 1< i \leq k\\
		\lambda_{i}^{(2)}= \lambda_{i+1}^{(2)} \times p_{i+1,i} & \mbox{if } 1 \leq i < k\\
		\lambda_{i}^{(2)}=\gamma_{i}   & \mbox{if } i=k
	\end{array}
\right.
\end{equation}

\subsubsection{Throughput}
The throughput can be calculated by adding the arrival rate of the second flow at $N_{1}$, and the arrival rate of the first flow at $N_{k}$ as follows
\begin{equation}
\label{eq:throughput2}
\theta= \lambda_{1}^{(2)} + \lambda_{k}^{(1)}
\end{equation}

\subsubsection{Successful transmission probabilities}
As explained earlier, to calculate the arrival rate at different nodes, we need to compute the probability of successful transmission in each link. The probability of successful transmission between different nodes, for the topology depicted in \figurename~\ref{chain}, can be calculated by finding a solution for the following system of non-linear equations
\begin{equation}
\label{eq:probabilities2-1}
\begin{cases} 
p_{1,2}=(1-2 \delta \lambda_{2})(1-2 \delta \lambda_{3}) \\ 
...\\
p_{i-1,i}=(1-2 \delta \lambda_{i})(1-2 \delta \lambda_{i+1}) \\
...\\
p_{k-2,k-1}=(1-2 \delta \lambda_{k-1})(1-2 \delta \lambda_{k}^{(2)})\\
p_{k-1,k}=(1-2 \delta \lambda_{k}^{(2)})\\
p_{k,k-1}=(1-2 \delta \lambda_{k-1})(1-2 \delta \lambda_{k-2}) \\
...\\
p_{i+1,i}=(1-2 \delta \lambda_{i})(1-2 \delta \lambda_{i-1})\\
...\\
p_{3,2}=(1-2 \delta \lambda_{2})(1-2 \delta \lambda_{1}^{(1)})\\
p_{2,1}=(1-2 \delta \lambda_{1}^{(1)})\\
\end{cases}
\end{equation}
where all $\lambda_{i}$s are functions of $\gamma_{1}$, $\gamma_{k}$, and successful transmission probabilities as described in~(\ref{eq:lambda2}).

\subsection{Step 3- One Flow, Retransmission, no Coding}\label{step3}
\begin{figure}[ht]
\centering
\includegraphics[scale=0.6]{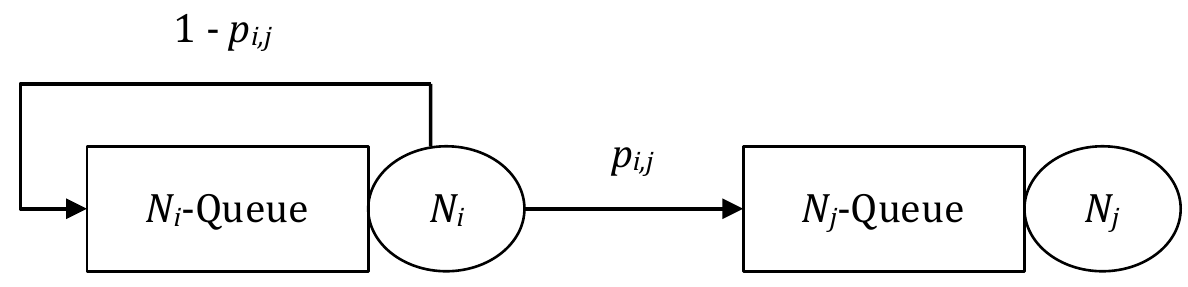}
\caption{Feedback queue to model retransmission.}
\label{feedback}
\end{figure}

To model retransmission of the packets in the network, feedback queues are required. As shown in \figurename~\ref{feedback}, we assume node $N_{i}$ delivers its packets to the next hop, $N_{j}$ with the probability $p_{i,j}$, and retransmits the packets with the probability $1-p_{i,j}$, at most $\beta$ times, (i.e., the packet is retransmitted if the last transmission fails). Note that we assume that the feedback channel is reliable, and ACK messages are received successfully. 

Taking retransmissions into account, the arrival rate of the queues increases as a portion of sent packets returns to the queue. Therefore, the arrival rate at nodes can be computed as
\begin{equation}
\label{eq:lambda3}
\lambda_{i} =
\left\{
	\begin{array}{ll}
		\gamma_{i}+\lambda_{i} (1-p_{i,i+1})   & \mbox{if } i=1\\
		\lambda_{i-1} p_{i-1,i} + \lambda_{i} (1-p_{i,i+1})  & \mbox{if } 1 < i < k\\
		\lambda_{i-1} p_{i-1,i}    & \mbox{if } i=k
	\end{array}
\right.
\end{equation}

The throughput equals the arrival rate at the destination $N_{k}$, denoted by $\lambda_{k}$. Also, the probability of successful transmission in different nodes is calculated by solving the system of non-linear equations presented in~(\ref{eq:probabilities1-1}).

\subsection{Step 4- Two Flows, Retransmission, no Coding}
The next step is adding the second flow to the case that nodes retransmit a packet if its transmission fails. 
In this scenario, the input rate from outside is non-zero at sources $N_{1}$ and $N_{k}$, known as $\gamma_{1}$ and $\gamma_{k}$, respectively, and each packet traverses the whole chain to be delivered at the other end of the chain.

As in Subsection~\ref{step2}, $\lambda_{i}^{(1)}$ denotes the arrival rate of the first flow at $N_{i}$, and $\lambda_{i}^{(2)}$ indicates the arrival rate of the second flow. Thus, the total arrival rate at each node equals $\lambda_{i}=\lambda_{i}^{(1)} + \lambda_{i}^{(2)}$, while as explained in Subsection~\ref{step3}, retransmissions should be taken into account in the calculation of $\lambda_{i}^{(1)}$ and $\lambda_{i}^{(2)}$. Eq.~(\ref{eq:lambda4}) presents the arrival rate of two flows at all nodes.

\begin{equation}
\label{eq:lambda4}
\left\{
	\begin{array}{ll}
	   \lambda_{i}^{(1)}=\gamma_{i} + \lambda_{i}^{(1)} (1-p_{i,i+1})  & \mbox{if } i=1\\
		\lambda_{i}^{(1)}= \lambda_{i-1}^{(1)} p_{i-1,i} + \lambda_{i}^{(1)} (1-p_{i, i+1}) & \mbox{if } 1< i < k\\
		\lambda_{i}^{(1)}= \lambda_{i-1}^{(1)} p_{i-1,i}  & \mbox{if } i=k\\
		\lambda_{i}^{(2)}=\gamma_{i} + \lambda_{i}^{(2)} (1-p_{i,i-1})  & \mbox{if } i=k\\
		\lambda_{i}^{(2)}= \lambda_{i+1}^{(2)} p_{i+1,i} + \lambda_{i}^{(2)} (1- p_{i, i-1}) & \mbox{if } 1 < i < k\\
		\lambda_{i}^{(2)}= \lambda_{i+1}^{(2)} p_{i+1,i} & \mbox{if } i=1
	\end{array}
\right.
\end{equation}

The throughput can be computed using~(\ref{eq:throughput2}), while~ the system of non-linear equations in~(\ref{eq:probabilities2-1}) is solved to find the probability of successful transmission in different links.

\subsection{Step 5 - Two Flows, Coding, no Retransmission}
To extend our model to the case that nodes can combine packets of two flows (i.e., the first flow from $N_{1}$ to $N_{k}$, and the second flow from $N_{k}$ to $N_{1}$), we need to distinguish native and coded packets from each other. In this model, we assume that the native and coded packets enter separate queues. Furthermore, coded packets in $Q_\text{coded}$ have a non-preemptive higher priority over the native packets in $Q_\text{native}$. This means that a coded packet will be forwarded earlier than all the packets waiting in the native queue, but a native packet in service (i.e., the native packet which is being transmitted) is not interrupted by coded packets.

As in the previous cases, we assume that the rate of generating packets at $N_{1}$ and $N_{k}$ equals $\gamma_{1}$ and $\gamma_{k}$, respectively, and $\lambda_{i}^{(1)}$ and $\lambda_{i}^{(2)}$ represent the rate of the first and the second flow at $N_{i}$, respectively. Also, we define $\lambda_{i}^{n}$ as the arrival rate of native packets, and $\lambda_{i}^{c}$ as the arrival rate of coded packets at $N_{i}$. 

\subsubsection{Coding module}
\begin{figure}[ht]
\centering
\includegraphics[scale=0.39]{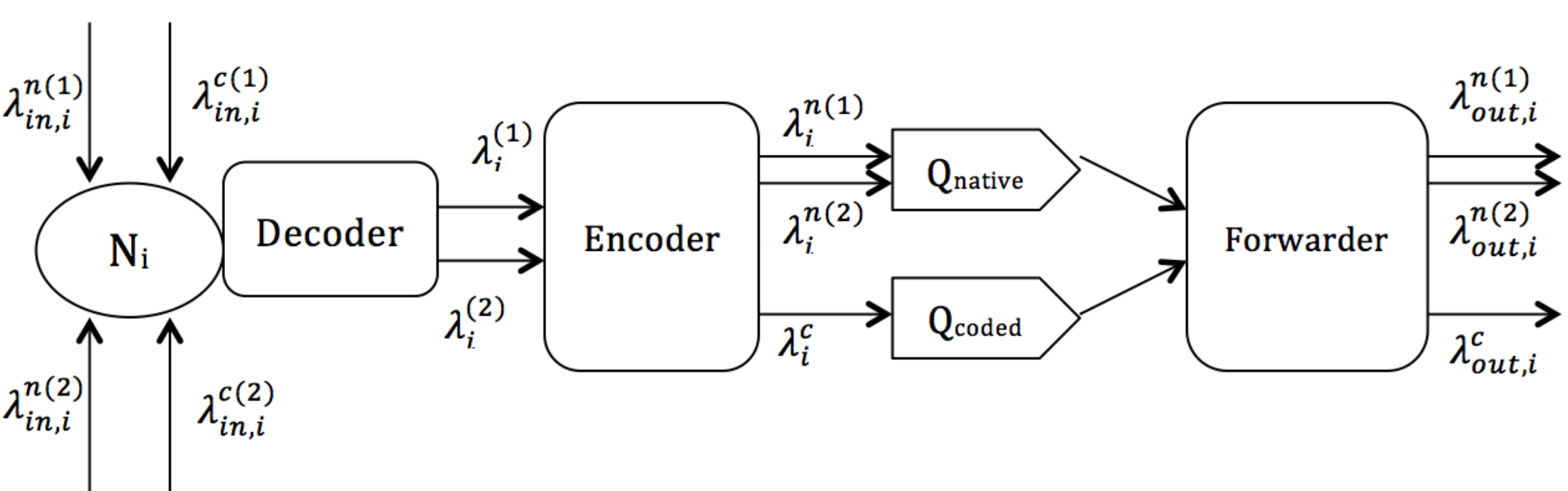}
\caption{A packet from arrival until departure.}
\label{coding-module}
\end{figure}

As shown in \figurename~\ref{coding-module}, $N_{i}$ receives native and coded packets of both flows from the previous hops. Although a coded packet is the combination of both flows, the receiver $N_{i}$ is the next hop of either the first flow or the second flow (i.e., intended flow). Due to this reason, we separate coded packets of different flows arriving at $N_{i}$.

The \emph{decoder}, in \figurename~\ref{coding-module}, checks the correctness of received packets, decodes the coded packets and finds the next hop of the packets. The outputs of this module are native packets of the first flow and the second flow with rates $\lambda_{i}^{(1)}$ and $\lambda_{i}^{(2)}$, respectively. In fact $\lambda_{i}^{(1)}$ ($\lambda_{i}^{(2)}$) equals the sum of the arrived native packets of the first (second) flow, denoted by $\lambda_{in,i}^{n(1)}$ ($\lambda_{in,i}^{n(2)}$) and the successfully decoded packets of the first (second) flow, represented by $\lambda_{in,i}^{c,(1)} \times p_{i-1,i}$($\lambda_{in,i}^{c,(2)} \times p_{i+1,i}$). Table~\ref{table:formula} shows the required equations to compute these variables as well as other described variables.

Note that if $N_{i}$ receives coded packet $P_{1} \oplus P_{2}$, and $P_{1}$ is its intended packet (i.e., the packet that this node was its next hop), to be able to decode $P_{1}$, $N_{i}$ should have already received $P_{2}$ from the other flow (i.e., from the opposite direction). Therefore, one could state $P_{1}$ will be received successfully if both $P_{1} \oplus P_{2}$  and $P_{2}$ have been received successfully. 

\begin{table}[!t]
\caption{The calculation of some variables' value.}
\label{table:formula}
\centering
\begin{tabular}{|c||c|}
\hline
\textbf{Variable} & \textbf{Equation}\\  \hline
$\lambda_{in,i}^{n(1)}$ &  $\lambda_{out,i-1}^{n(1)} \times p_{i-1,i}$ \\ \hline
$\lambda_{in,i}^{n(2)}$ &  $\lambda_{out,i+1}^{n(2)} \times p_{i+1,i}$	\\ \hline
$\lambda_{in,i}^{c(1)}$ &  $\lambda_{out,i-1}^{c} \times p_{i-1,i}$\\ \hline
$\lambda_{in,i}^{c(2)}$ &  $\lambda_{out,i+1}^{c} \times p_{i+1,i}$\\ \hline
$\lambda_{i}^{(1)}$ &  $\lambda_{in,i}^{n(1)} + \lambda_{in,i}^{c(1)} \times p_{i+1,i}$\\ \hline
$\lambda_{i}^{(2)}$ &  $\lambda_{in,i}^{n(2)} + \lambda_{in,i}^{c(2)} \times p_{i-1,i}$	\\ \hline
$\lambda_{i}^{c}$ &  $\min(\lambda_{i}^{(1)} , \lambda_{i}^{(2)}) \times P_\text{mix}$	\\ \hline
$\lambda_{i}^{n(1)}$ & $\lambda_{i}^{(1)} - \min(\lambda_{i}^{(1)} , \lambda_{i}^{(2)}) \times P_\text{mix}$. \\ \hline
$\lambda_{i}^{n(2)}$ & $\lambda_{i}^{(2)} - \min(\lambda_{i}^{(1)} , \lambda_{i}^{(2)}) \times P_\text{mix}$. \\ \hline
\end{tabular}
\end{table}

Since we do not delay the transmission of native packets in favor of generating more coded packets, a packet may be transmitted natively if it is at the head of $Q_\text{native}$ and there is no packet in $Q_\text{coded}$. Therefore, the \emph{encoder} receives the arrived native packet $P$ and looks for a packet from the other flow in $Q_\text{native}$. If the node can find such a packet $P\prime$, it will remove $P\prime$ from $Q_\text{native}$, mix it with $P$ and add the coded packet to $Q_\text{coded}$; otherwise, it will add $P$ to $Q_\text{native}$. Hence, the arrival rate of coded packets in $Q_\text{coded}$ (i.e., $\lambda_{i}^{c}$) is equal to the minimum arrival rate of both native flows ($\min(\lambda_{i}^{(1)} , \lambda_{i}^{(2)})$) multiplied by $P_\text{mix}$, where $P_\text{mix}$ is the probability of mixing the packets of two flows. The rest of the packets arrive in $Q_\text{native}$, and are transmitted natively. Therefore, $\lambda_{i}^{n(1)} = \lambda_{i}^{(1)} - \min(\lambda_{i}^{(1)} , \lambda_{i}^{(2)}) \times P_\text{mix}$. Also,  $\lambda_{i}^{n(2)}$ can be calculated in a similar way, as presented in Table~\ref{table:formula}.

\subsubsection{Native and coded queues}
The arrival rates in $Q_\text{native}$ and $Q_\text{coded}$ equal $\lambda_{i}^{n(1)}+\lambda_{i}^{n(2)}$ and $\lambda_{i}^{c}$, respectively. The \emph{forwarder} module, in \figurename~\ref{coding-module}, is responsible for forwarding packets. If $Q_\text{coded}$ is not empty, it will select the packet from the head of $Q_\text{coded}$; otherwise, the packet is chosen from the head of $Q_\text{native}$ if $Q_\text{native}$ is not empty.

As stated earlier, priority queues are used to model this case, where the arrival rate in $Q_\text{native}$ is the sum of the arrival rate of both flows (i.e., $\lambda_{i}^{n}=\lambda_{i}^{n(1)}+\lambda_{i}^{n(2)}$), and the total arrival rate in the queuing system of $N_{i}$ is presented by $\lambda_{i}=\lambda_{i}^{n}+\lambda_{i}^{c}$. 

By knowing the input rate of native and coded packets at all nodes, one can calculate the output rate in different nodes. Note that since we assume the queuing system is in a stable state, the departure rates equal the arrival rates ($\lambda_{out,i}^{n(1)}=\lambda_{i}^{n(1)}, \lambda_{out,i}^{n(2)}=\lambda_{i}^{n(2)}, \lambda_{out,i}^{c}=\lambda_{i}^{c}$). Finally, as stated in previous sections, the throughput equals the input rate of the second flow at $N_{1}$ plus the input rate of the first flow at $N_{k}$ (i.e., $\lambda_{1}^{(2)}+\lambda_{k}^{(1)}$). 

\begin{table}[!t]
\caption{Input rate of native packets at all nodes.}
\label{table:inputRateNative}
\centering
\begin{tabular}{|c|c|c|}
\hline
\textbf{$i$}&\textbf{$\lambda_{in,i}^{n(1)}$} & \textbf{$\lambda_{in,i}^{n(2)}$}\\  \hline
$i=1$ & $\gamma_{1}$ & $\lambda_{out,i+1}^{n(2)} \times p_{i+1,i}$ \\ \hline
$ 1 < i < k$ & $\lambda_{out,i-1}^{n(1)} \times p_{i-1,i}$ & $\lambda_{out,i+1}^{n(2)} \times p_{i+1,i}$\\ \hline
$i=k$ & $\lambda_{out,i-1}^{n(1)} \times p_{i-1,i}$ & $\gamma_{k}$\\ \hline
 \end{tabular}
\end{table}

\begin{table}[!t]
\caption{Input rate of coded packets at all nodes.}
\label{table:inputRateCoded}
\centering
\begin{tabular}{|c|c|c|}
\hline
\textbf{$i$}& \textbf{$\lambda_{in,i}^{c(1)}$} & \textbf{$\lambda_{in,i}^{c(2)}$}\\  \hline
$i=1, i=2$ & $0$ & $\lambda_{out,i+1}^{c} \times p_{i+1,i}$\\ \hline
$ 2 < i < k-1$ & $\lambda_{out,i-1}^{c} \times p_{i-1,i}$ & $\lambda_{out,i+1}^{c} \times p_{i+1,i}$\\ \hline
$i=k, i=k-1$ & $\lambda_{out,i-1}^{c} \times p_{i-1,i}$ & $0$\\ \hline
 \end{tabular}
\end{table}

Tables~\ref{table:inputRateNative} and \ref{table:inputRateCoded} provide the input rate of native and coded packets at all nodes. Moreover, it is clear that the output rate of the first flow at $N_{1}$ and the output rate of the second flow at $N_{k}$ equal $\gamma_{1}$ and $\gamma_{k}$, respectively. In addition, the output rate of the second flow and coded packets at $N_{1}$ and the output rate of the first flow and coded packets at $N_{k}$ are equal to zero, as presented in~(\ref{eq:lambda-out}).
\begin{equation}
\label{eq:lambda-out}
\left\{
	\begin{array}{ll}
	   \lambda_{out,1}^{n(1)}=\gamma_{1} \\
		\lambda_{out,k}^{n(2)}=\gamma_{k} \\
		\lambda_{out,1}^{n(2)}= 0 \\
		\lambda_{out,k}^{n(1)}=0 \\
		\lambda_{out,i}^{c}= 0& \mbox{if } i=1,k
	\end{array}
\right.
\end{equation}

\subsection{Step 6 - Two Flows, Coding, Retransmission}
In this scenario, we assume if a packet is not received successfully by the next hop, it will be retransmitted at most $\beta$ times. Such a packet arrives at the $Q_\text{coded}$ if there exists any coding opportunity; otherwise, it is inserted at the head of $Q_\text{native}$. Therefore, the arrival rates at the \emph{encoder} ($\lambda_{i}^{(1)}$ and $\lambda_{i}^{(2)}$), in \figurename~\ref{coding-module}, should be recalculated as follows
\begin{equation}
\left\{
	\begin{array}{ll}
	   \lambda_{i}^{(1)} = \lambda_{in,i}^{n(1)} + \lambda_{in,i}^{c(1)} (1-(1-p_{ i+1,i})^\beta)+ \lambda_{i}^{(1)} (1- p_{i, i+1})     \\ ~~~~~~~~~~~~~~~~~~~~~~~~~~~~~~~~~~~~~~~~~~~ \mbox{if } 1 \leq i < k\\
	   \lambda_{i}^{(1)} = \lambda_{in,i}^{n(1)} + \lambda_{in,i}^{c(1)}    ~~~~~~~~~~~~~~~~~~~ \mbox{if } i= k\\
	   \lambda_{i}^{(2)} = \lambda_{in,i}^{n(2)} + \lambda_{in,i}^{c(2)}      ~~~~~~~~~~~~~~~~~~~\mbox{if } i=1 \\
		\lambda_{i}^{(2)} = \lambda_{in,i}^{n(2)} + \lambda_{in,i}^{c(2)} (1-(1-p_{ i-1,i})^\beta) + \lambda_{i}^{(2)}  (1-p_{i, i-1})     \\~~~~~~~~~~~~~~~~~~~~~~~~~~~~~~~~~~~~~~~~~~~\mbox{if } 1 < i \leq k
	\end{array}
\right.
\end{equation}
Then, the throughput can be computed using (\ref{eq:throughput2}).

\section{Performance Evaluation} \label{performanceEvaluation}
To verify the accuracy of our proposed model, we run simulations in NS-2 for the topology depicted in \figurename~\ref{chain-5}, and compare the simulation and analytical results. As shown in this figure, $N_{1}$ and $N_{5}$ transmit their packets to each other via intermediate nodes $N_{2}$, $N_{3}$ and $N_{4}$.

\subsection{Network Description}
In our simulation, we use the IEEE 802.11 standard as the MAC layer protocol, and a node may drop a packet due to collision. Based on the specifications, a node transmits a packet at most $7$ times (i.e., $\beta=7$). The link rate is set to 2 Mbps. The sources, in our simulation scenarios, send packets according to a Poisson process with a datagram size of $1000$ bytes. Also, we use Destination-Sequenced Distance-Vector Routing (DSDV) as the routing protocol, and we assume $P_\text{mix}=0.5$. We compare the analytical result with the simulation results in different scenarios in terms of throughput in a chain topology with $5$ nodes. 

\subsection{Evaluation Results}
The comparison of simulation and analytical results of the topology depicted in \figurename~\ref{chain-5} are shown in Tables~\ref{table:step12-th}-\ref{table:step56-th}. In our simulations, the flows between $N_{1}$ and $N_{5}$ last for 170 seconds. We change the arrival rate of packets at sources, and calculate the total throughput. In case of two flows, we assume that the arrival rates at sources are equal (i.e., $\gamma_{1}=\gamma_{k}$).


\begin{table}[!t]
\caption{Throughput comparison for no coding cases without retransmission.}
\label{table:step12-th}
\centering
\begin{tabular}{|c|c|c|c|}
\hline
\textbf{Arrival} & \textbf{Number} &  \textbf{Analysis} & \textbf{Simulation} \\ 
\textbf{rate} & \textbf{of flows} &   &  \\  \hline
10& 1 &	 9.37	 & 9.31 \\ \hline
14.286 & 1 & 13.39 &	13.29\\ \hline
20 & 1 & 18.74	& 18.63\\ \hline
25 & 1&  23.43 &  23.3 \\ \hline
10 &  2 & 18.73	 & 18.67\\ \hline
14.286 & 2 &  26.77 &	26.62 \\ \hline
20 & 2 & 37.46	 & 37.29 \\ \hline
25 & 2 & 46.82   & 46.23 \\ \hline
\end{tabular}
\end{table}

\begin{table}[!t]
\caption{Throughput comparison for no coding cases with retransmission.}
\label{table:step34-th}
\centering
\begin{tabular}{|c|c|c|c|}
\hline
\textbf{Arrival} & \textbf{Number} & \textbf{Analysis} & \textbf{Simulation} \\ 
\textbf{rate} & \textbf{of flows} &   &  \\  \hline
10& 1 &	 10	& 10.02 \\ \hline
14.286 & 1 & 14.29 & 14.31 \\ \hline
20 & 1 &  20	& 20.02 \\ \hline
25 & 1 &  25  & 25.01 \\ \hline
10 &  2 &   20	 & 20.01 \\ \hline
14.286 & 2 & 28.57	& 28.56\\ \hline
20 & 2 & 40 & 39.94\\ \hline
25 & 2 &  50 & 49.59 \\ \hline
\end{tabular}
\end{table}

\begin{table}[!t]
\caption{Throughput comparison for coding cases.}
\label{table:step56-th}
\centering
\begin{tabular}{|c|c|c|c|}
\hline
\textbf{Arrival} & \textbf{Retransmission} & \textbf{Analysis} & \textbf{Simulation} \\ 
\textbf{rate} & && \\ \hline
10& 	no & 18.45 &	18.68\\ \hline
14.286  & no & 26.35	 & 26.68\\ \hline
20 & no & 36.89	& 37.29\\ \hline
25 & no & 46.62	& 46.36\\ \hline
10& 	 yes & 20	 & 20.02\\ \hline
14.286 &	yes & 28.57	 & 	28.59  \\ \hline
20 & yes &  40 &	40.035 \\ \hline
25 & yes & 49.98	&	49.64 \\ \hline
\end{tabular}
\end{table}

Tables~\ref{table:step12-th}, \ref{table:step34-th} and~\ref{table:step56-th} present the throughput for different steps explained in this paper. As shown in these tables, the simulation results of throughput closely match the analytical results  in different arrival rates for both the traditional non-coding scheme and the coding scheme. This consistency of the simulation and analytical results corroborates the validity of our analytical model proposed in previous section. 

Note that here, we compare the analytical results derived from our model with simulation results. We do not aim to compare the performance of coding and non-coding schemes. This is due to the fact that without delaying native packets, network coding usually shows its gain over the traditional forwarding approach where arrival rates are high enough to provide frequent coding opportunities. However, at higher arrival rates, it is difficult to hold the assumption of having stable queues in our simulation scenarios, especially for non-coding scheme.

\balance
\section{Conclusion and Future Work} \label{conclusion}
In this paper, queueing theory was applied to derive the throughput of inter-flow network coding in multi-hop wireless mesh networks, where two unicast sessions in opposite directions traverse the network. We proposed an analytical framework considering the specifications of the IEEE 802.11 standard to formulate the collision probability of links, and retransmissions. Our analytical model assumes M/M/1 queues, which are in a stable state, while coded and native packets arrive at separated queues and coded packets have a non-preemptive higher priority over native packets. We verified the accuracy of the proposed analytical model by computer simulation in NS-2.

This worth noting that although we provided our model for a chain topology with two flows, our formulation of collision can be applied to any topology. A future extension of our work will be to present an analytical framework for a general topology, where more than two flows are traveling and possibly mixing together. In addition, we plan to add cooperative forwarding to our model, where the neighbors of the next-hop could forward the packet if the next-hop does not receive it. 

\bibliographystyle{IEEEtran} 
\bibliography{citation} 

\end{document}